# Auto-ICell: An Accessible and Cost-Effective Integrative Droplet Microfluidic System for Real-Time Single-Cell Morphological and Apoptotic Analysis

Yuanyuan Wei[1], Meiai Lin[2], Shanhang Luo[2,3], Syed Muhammad Tariq Abbasi[1], Liwei Tan[4], Guangyao Cheng[1], Bijie Bai[5], Yi-Ping Ho[1,6,7,8], Scott Wu Yuan[1,*], Ho-Pui Ho[1,*]

[1] Department of Biomedical Engineering, The Chinese University of Hong Kong, Shatin, Hong Kong SAR, China. E-mail: wyuan@cuhk.edu.hk; aaron.ho@cuhk.edu.hk

[2] Department of Biomedical Engineering, Shantou University, Shantou, China

[3] Department of Biomedical Engineering, National University of Singapore, Singapore

[4] Department of Electrical Engineering, Tsinghua University, Beijing, China

[5] Electrical and Computer Engineering Department, University of California, Los Angeles, California, 90095, USA

[6] Centre for Biomaterials, The Chinese University of Hong Kong, Hong Kong SAR, China

[7] Hong Kong Branch of CAS Center for Excellence in Animal Evolution and Genetics, Hong Kong SAR, China

[8] State Key Laboratory of Marine Pollution, City University of Hong Kong, Hong Kong SAR, China

[*]Correspondence: wyuan@cuhk.edu.hk; aaron.ho@cuhk.edu.hk.

**Abstract**

The Auto-ICell system, a novel, and cost-effective integrated droplet microfluidic system, is introduced for real-time analysis of single-cell morphology and apoptosis. This system integrates a 3D-printed microfluidic chip with image analysis algorithms, enabling the generation of uniform droplet reactors and immediate image analysis. The system employs a color-based image analysis algorithm in the bright field for droplet content analysis. Meanwhile, in the fluorescence field, cell apoptosis is quantitatively measured through a combination of deep-learning-enabled multiple fluorescent channel analysis and a live/dead cell stain kit. Breast cancer cells are encapsulated within uniform droplets, with diameters ranging from 70 μm to 240 μm, generated at a high throughput of 1,500 droplets per minute. Real-time image analysis results are displayed within 2 seconds on a custom graphical user interface (GUI). The system provides an automatic calculation of the distribution and ratio of encapsulated dyes in the bright field, and in the fluorescent field, cell blebbing and cell circularity are observed and quantified respectively. The Auto-ICell system is non-invasive and provides online detection, offering a robust, time-efficient, user-friendly, and cost-effective solution for single-cell analysis. It significantly enhances the detection throughput of droplet single-cell analysis by reducing setup costs and improving operational performance. This study highlights the potential of the Auto-ICell system in advancing biological research and personalized disease treatment, with promising applications in cell culture, biochemical microreactors, drug carriers, cell-based assays, synthetic biology, and point-of-care diagnostics.

## 1. Introduction

Cell heterogeneity, inherent even in cells from the same microenvironment, necessitates single-cell analysis to reveal mechanisms obscured by cell bulk population averages[1]. Single-cell cultivation is crucial for periodic individual cell investigation, promoting the growth of single-cell colonies and single-cell-derived secondary metabolites, thereby facilitating further single-cell level analysis[2,3]. Periodic and sequential single-cell observation is key to various cellular and subcellular studies, including morphology, proliferation, differentiation, migration, omics, and epigenomics. Droplet microfluidics presents an efficient, sensitive method for encapsulating single cells into uniform compartments, enabling real-time analysis both on-chip and off-chip[3–7]. This technique promotes cellular and subcellular investigations for diagnosis and therapy. It is also utilized in compartmentalizing biochemical reactions, shielding them from external interference, and enabling applications in single-cell analysis, kinetic studies, and controlled drug release[8,9].

However, the development of droplet microfluidics for single-cell analysis faces challenges from both set-up and detection scheme perspectives[10]. Conventional chip fabrication methods rely heavily on expensive instruments, sophisticated operations, and clean-room conditions[11,12]. Limitations of Polydimethylsiloxane (PDMS), including its incompatibility with large-scale manufacturing and packaging, and difficulty in creating complex 3D structures, further restrict design optimization[13–15]. To overcome these hurdles, scientists have explored alternative fabrication methods and materials. Laser-based fabrication, for instance, enables fast prototyping of complex 3D structures but still requires sophisticated instrumentation and manual operations[–19]. Additive manufacturing, or 3D printing, has emerged as a cost-effective solution for microfabrication[20–22]. Its capability for applying to microfluidics has been validated, offering simplicity, the ability to create complex structures, and suitability for large-scale production.

The observation of cellular phenomena in individual cells also heavily relies on expensive facilities and expert operators[10]. Tracking effort therefore increases dramatically for each additional generation. The biggest bottleneck in data analysis therefore is the time needed for cell tracking; users typically spend days to weeks manually extracting and plotting data[23]. The conventional detection schemes for analyzing fluorescence images of stationary cells rely on software tools with manual threshold detection, such as Image J, Fiji, and Cell Profiler[23,24]. However, this approach can be labor-intensive and lacks scalability. Alternative approaches like in-flow interrogation[9], Raman spectroscopy[25], mass spectrometry[26], electrical impedance spectroscopy, and electrophysiological recording, often require expensive and complex devices[7], limiting speed, portability, and operational simplicity. For automatic and sensitive analysis, computational image analysis and deep learning have emerged as powerful tools in biosensing and data analysis, including droplet microfluidics. However, the application at this stage is only limited to physical mechanisms (e.g., coalescence) and droplet sorting [27–30]. Therefore, there is an urgent need for dedicated and integrative microfluidic devices that fulfill the ideal conditions for culturing and analyzing individual cells, such as being label-free, non-invasive, high-efficiency, high-throughput, and suitable for long-term culture.

Here, we report an accessible and cost-effective integrative droplet microfluidic system, Auto-ICell, for real-time single-cell morphological and apoptotic analysis. This system combines a 3D-printed microfluidic chip for generating uniform droplet reactors, with image analysis algorithms for real-time image analysis. In the bright field, a color-based image analysis algorithm enables droplet content analysis. In the fluorescence field, cell apoptosis is measured quantitatively by combining deep-learning-enabled multiple fluorescent channel analysis with a live/dead cell stain kit.

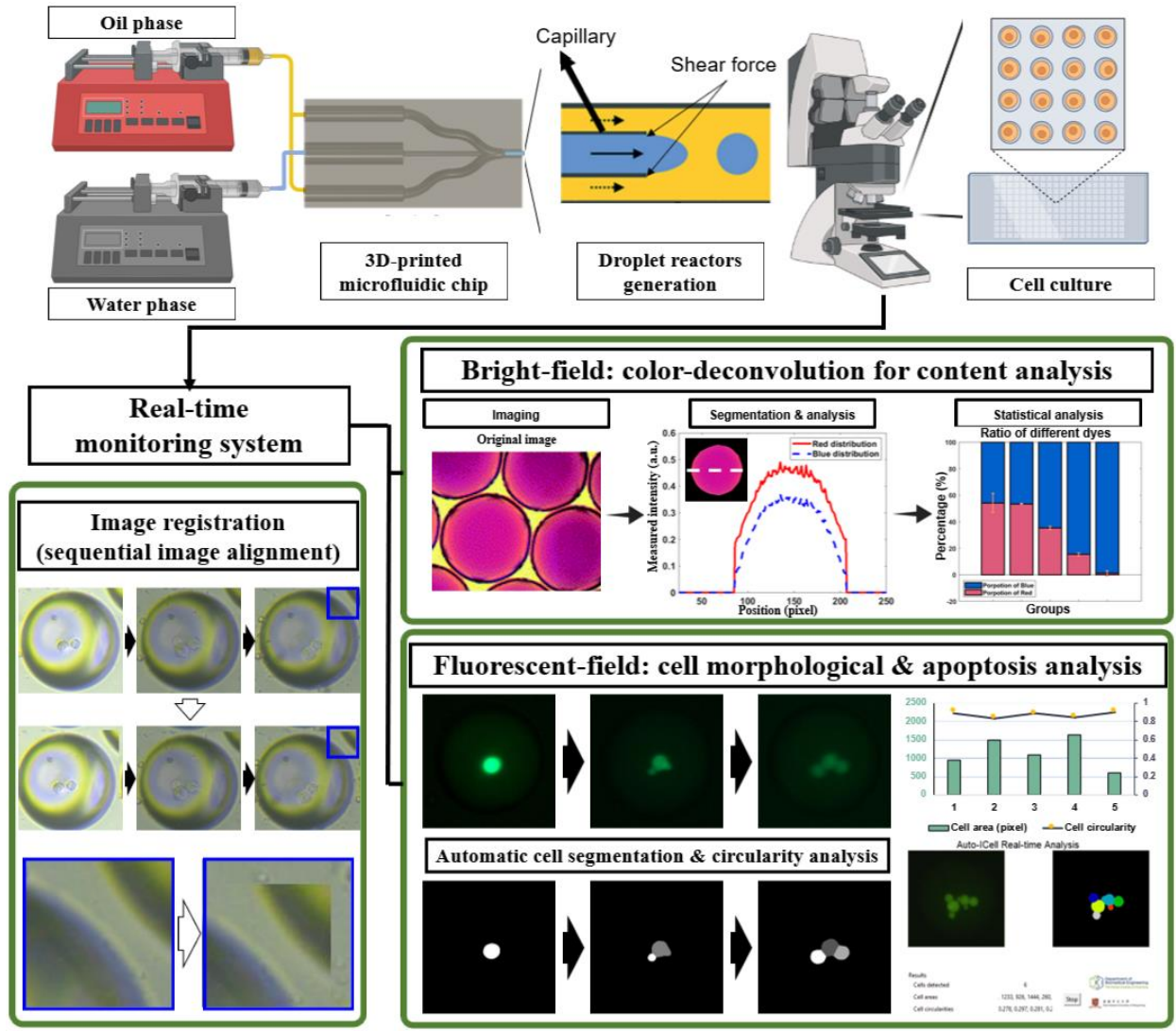

**Fig. 1. Illustration of the Auto-ICell pipeline for real-time, single-cell morphological and apoptotic analysis.** The pipeline integrates droplet reactor generation and real-time image analysis, leveraging both bright-field and fluorescence-field modalities. Breast cancer cells are encapsulated within uniform droplets, with diameters ranging from 70 μm to 240 μm, produced using a 3D-printed microfluidic chip and a customized microfluidic system. By optimizing the syringe pumps connected to the microfluidic chip, we achieve a generation throughput of 1,500 droplets per minute. The resulting droplet reactors are collected in a 96-microwell cell culture plate for imaging and subsequent analysis. Images are captured using a laboratory fluorescent microscope and the analysis results are displayed in real-time, within 2 seconds, on a custom graphical user interface (GUI). In the bright-field, an image analysis algorithm based on color deconvolution automatically

calculates the distribution and ratio of encapsulated dyes. In the fluorescence field, cell blebbing is observed and cell circularity is quantified using a deep-learning-enabled cell segmentation algorithm. To enhance accuracy and mitigate the impact of reactor shifting caused by factors such as thermal fluctuations, mechanical vibrations, and Brownian motion, sequential images are aligned using image registration before analysis. The non-invasive Auto-ICell system provides online detection, offering a robust, time-efficient, user-friendly, and cost-effective solution.

2. **Results**

Our integrative Auto-ICell system was validated through the encapsulation and monitoring of MDA-MB-231 breast cancer cells, exemplifying long-term 3D cell culture and biochemical reactions. As depicted in **Fig. 1**, the process encompasses the generation of droplet reactors and real-time image analysis in both bright-field and fluorescence-field settings. The pipeline initiates with the encapsulation of cells within uniform droplet reactors, created using a 3D-printed microfluidic chip and a custom microfluidic system. The chip is fabricated using a stereolithography 3D printer (resolution: 10 μm) with photosensitive resin as the printing material. The system achieves a high throughput of 1,500 droplets per minute, generating droplets with a programmable diameter ranging from 70 μm to 240 μm. This is achieved by fine-tuning the input of the water phase and oil phase (mineral oil with 1% span80 surfactant). The resulting droplet reactors are collected into a 96-microwell cell culture plate for cell culture, staining, imaging, and subsequent analysis. Images of these droplet reactors are captured by a laboratory fluorescent microscope, with the corresponding analysis results displayed on a custom graphical user interface (GUI) in real time, typically within 2 seconds. In the bright-field, a color-deconvolution-based image analysis algorithm automatically detects and visualizes the distribution and ratio of encapsulated dyes. Simultaneously, in the fluorescence field, cell blebbing is observed and cell circularity is quantified using a deep-learning-enabled cell segmentation algorithm. To ensure process accuracy and to lessen the impact of reactor movement - caused by the temperature difference between the cell plate and the moving stage - image registration is performed to align sequential images prior to image analysis. This precise alignment facilitates a more detailed and accurate analysis of cell morphology and apoptosis.

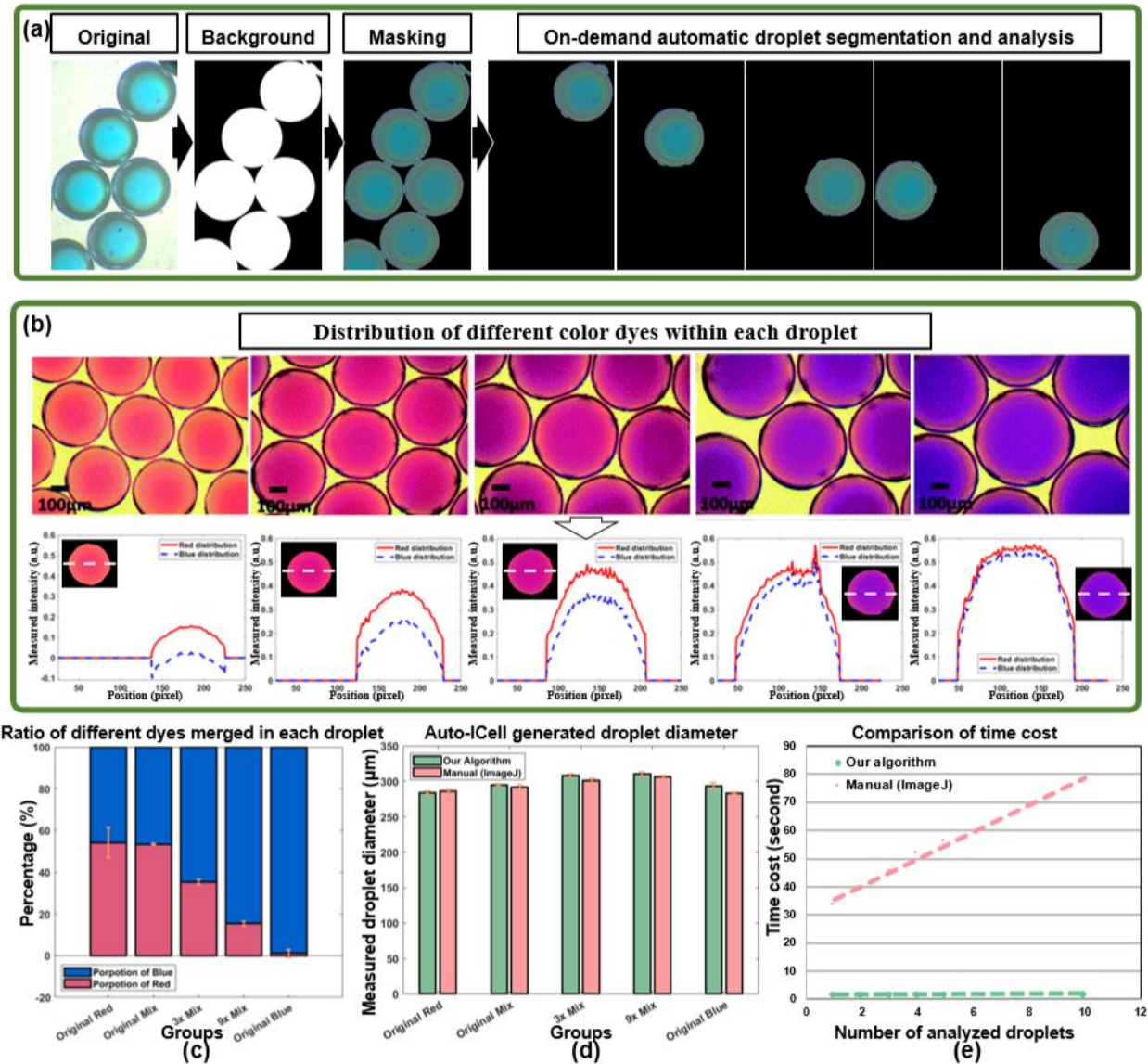

**Fig. 2 High-throughput analysis of encapsulated dye distribution and ratio using the Auto-ICell bright-field color-deconvolution algorithm.** (a) The algorithm workflow and performance are depicted, with droplets segmented from the entire field of view (FOV) image through a process involving background identification and binary masking. (b) The color-deconvolution process is employed to detect and visualize the intensity of specific target colors, allowing for the plotting of dye distributions within each droplet (Scale bar: 100 μm). (c) The algorithm automatically detects the ratio of encapsulated dyes in each droplet, thereby increasing the detection throughput of droplet reactors without the need for additional equipment. (d) The algorithm's droplet size measurements are compared with manual measurements using ImageJ software, showing comparable results. (e) Execution time comparison between our algorithm and the manual method demonstrates the significant reduction in operational execution time achieved by our accessible

and automatic algorithm. As the number of analyzed droplets increases, our algorithm maintains a time of less than 5 seconds, while the manual solution using ImageJ exhibits a linear increase from more than 30 seconds.

**Fig. 2** presents the bright-field color-deconvolution algorithm incorporated in the Auto-ICell system, designed for high-throughput analysis of encapsulated dyes. The workflow and performance of this algorithm are illustrated in **Fig. 2(a)**. The process commences with the segmentation of droplets from the entire field of view (FOV) image. This step involves identifying the background and generating a binary mask, resulting in the successful on-demand segmentation of droplets within a single second. The workflow then proceeds with a color-deconvolution process to detect and visualize the intensity associated with specific target colors. For accuracy, the RGB value of target colors was obtained from droplets encapsulating a single dye, respectively. Consequently, the dye distributions within each droplet and the ratios of dyes are correspondingly plotted. This feature enhances the detection throughput of droplet reactors, negating the need for supplementary auxiliary equipment. We validated the algorithm by analyzing droplets encapsulating serially diluted mixed dyes, produced by our Auto-ICell system. The results are displayed in **Fig. 2(c)**. As the concentration of red dye decreased, the ratio of detected red dye diminished from 54.20% to 1.16%. The algorithm's generality was tested by analyzing droplets generated using various platforms and different dyes[31], with representative plotting results shown in **Fig. 2(b)**. After calibration, the algorithm successfully analyzed dye distributions of the serial droplets, without necessitating transfer learning, modifications, or retraining. Additional information, including the diameter and coefficient variation (C.V.) of droplets, is also simultaneously calculated. **Fig. 2(d)** provides a comparison between droplet size measurements obtained using the Auto-ICell algorithm and manual measurements taken with ImageJ software. The comparable results from both methods attest to the accuracy of the Auto-ICell algorithm. The algorithm's efficiency is underscored in **Fig. 2(e)**. A comparison of the execution time between the Auto-ICell algorithm and the manual method demonstrates a significant reduction in operation time when using the automated algorithm. As the number of droplets analyzed increases from 1 to 10, the Auto-ICell algorithm consistently processes the data in under 3 seconds. Conversely, the manual method using ImageJ software displays a linear increase in processing time, starting from over 30 seconds. This comparison underscores the efficiency and accessibility of the Auto-ICell algorithm, particularly for high-throughput applications.

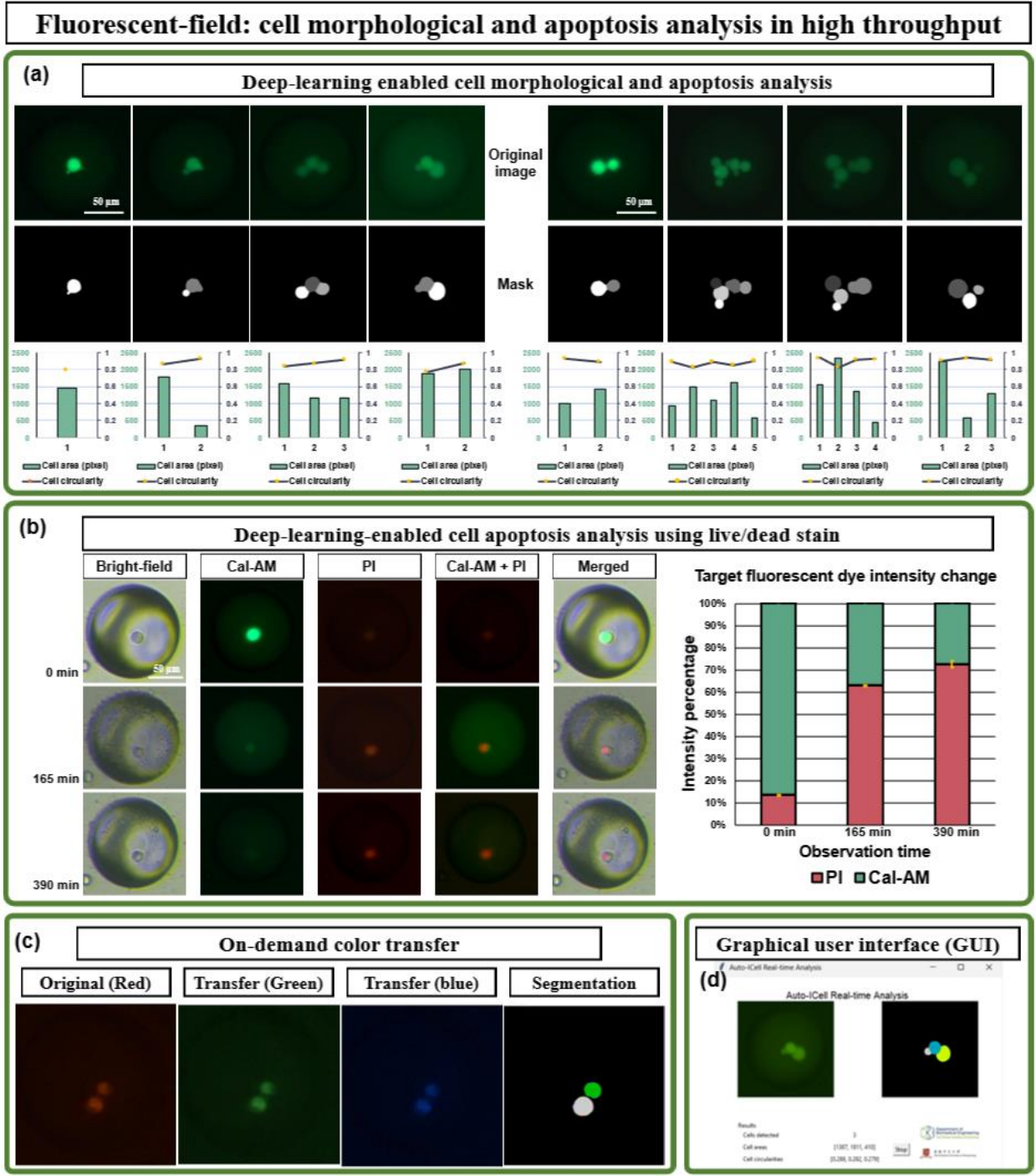

**Fig. 3 High-throughput cell morphological and apoptotic analysis using the Auto-ICell fluorescence field, deep-learning-enabled cell segmentation, and circularity calculation.** (a) The algorithm's workflow and performance are illustrated. Cells captured in fluorescence fields are automatically segmented from the entire FOV image using a deep-learning model. Parameters of the target cells, including size (area), perimeter, and circularity, are extracted and calculated automatically within 1 second. Cell morphology is characterized by circularity, while apoptosis is characterized by cell area. The process of breast cancer cell blebbing, along with the corresponding

masks and plotting of area and circularity, is presented (Scale bar: 50 μm). (b) Apoptosis is automatically monitored by applying a live/dead stain dye and quantifying the corresponding fluorescent color intensity. This kit enables simultaneous staining of living and dead cells using two fluorescent dyes. Living cells produce green fluorescence due to the conversion of Calcein acetoxymethyl ester (Calcein-AM) to Calcein, while dead cells produce red fluorescence due to the presence of propidium iodide (PI). Cell apoptosis can be quantified by calculating the ratio of the two fluorescence intensities. This approach provides a highly accurate and reliable method for observing cell membrane protrusions, capturing the life cycle of bleb initiation, expansion, and retraction. (c) We have integrated on-demand fluorescence color transfer technology into our algorithm, enhancing biological experiments through multiplexing and compatibility with existing assays. This technology facilitates the transfer of fluorescence color (e.g., from red to green and blue), offering flexibility in experimental design by allowing researchers to switch between different fluorescent colors without additional changes, eliminating the need for labeling, sample preparation steps, or expensive filter changes and modifications. (d) Images captured from the fluorescent microscope are displayed in real-time on our GUI, with cell segmentation results shown on the right side. Parameters, including cell counts, area, and circularity, are displayed correspondingly. This standardized and user-friendly GUI streamlines the workflow, visualizes results within 2 seconds, and saves the results in a designated folder.

In the realm of fluorescence-field imaging, Auto-ICell introduces a deep-learning-enabled cell segmentation and circularity calculation algorithm for high-throughput analysis of cell morphology and apoptosis. The algorithm's workflow and performance are demonstrated in **Fig. 3(a)**. Parameter tuning, typically reliant on experience and time-consuming, would ideally be automated[31]. Deep learning presents a path to this automation, as neural networks can autonomously learn the correlation between input data and the desired output, given sufficient training data. Using the deep-learning algorithm modified from an open-sourced tool, Cellpose[32], cells captured in fluorescence fields were automatically segmented from the entire FOV image. Based on the segmented mask, parameters such as size (area), perimeter, and circularity of the target cells were automatically extracted and calculated within a mere second. Cell morphology was characterized by cell circularity according to equation 1, while cell area was utilized to characterize cell apoptosis. Our algorithm was applied to study the process of breast cancer cell blebbing, presenting corresponding masks and plotting of area and circularity. To monitor cell apoptosis, we employed a live/dead stain dye and quantified the corresponding fluorescent color intensity, thereby enabling automatic monitoring of cell apoptosis. Our kit facilitated the simultaneous staining of living and dead cells using two fluorescent dyes. Living cells produced green fluorescence due to the conversion of Calcein acetoxymethyl ester (Calcein-AM) to Calcein, while dead cells emitted red fluorescence due to the presence of propidium iodide (PI). Cell apoptosis can be quantified by calculating the ratio of the two fluorescence intensities. This methodology provided a highly accurate and reliable technique for observing cell membrane protrusions, including the life cycle of bleb initiation, expansion, and retraction as depicted in **Fig. 3(b)**. Additionally, we integrated an on-demand fluorescence color transfer technology into our algorithm, as shown in **Fig. 3(c)**. This technology allows for the transfer of fluorescent colors from one to another (e.g., from red to green and blue), providing greater flexibility in experimental design. Researchers can easily switch between different fluorescent colors without the need for additional labeling, sample preparation steps, or costly filter changes and modifications. To facilitate the use of our algorithm, we developed a user-friendly GUI. **Fig. 3(d)** displays real-time images captured from the fluorescent microscope on our GUI, with cell segmentation results displayed on the right side. Parameters including cell counts, area, and circularity are displayed correspondingly. This standardized and user-friendly GUI streamlined the workflow, allowing for visualization of results within 2 seconds, and facilitated the documentation of results in a directed folder.

$$Circularity = 4\pi \times \frac{Area}{Perimeter^2} \tag{1}$$

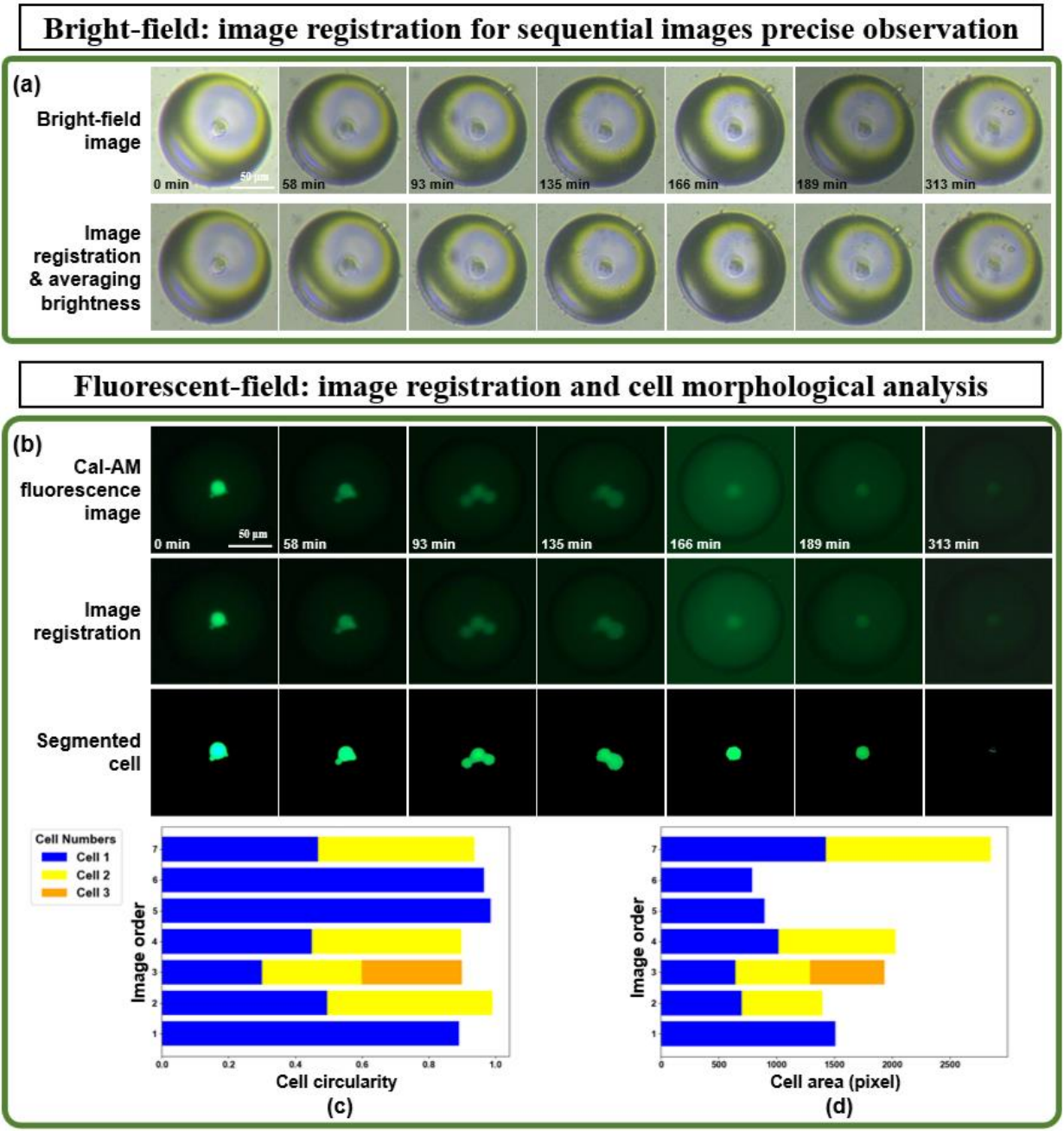

**Fig. 4 Auto-ICell image registration algorithm applied in bright-field and fluorescence-field for precise single-cell morphological and apoptotic analysis.** (a) In the bright field, sequential images of single cells encapsulated in droplets are aligned using an image registration algorithm to counteract the interference from shifting, which can occur due to thermal fluctuations, mechanical vibrations, Brownian motion, etc. The captured images are automatically aligned with the first image serving as the reference point, all within 2 seconds. (b) In the fluorescence field, the alignment of sequential images of single cells encapsulated in droplets is performed similarly. Additionally, the algorithm incorporates other analysis functions, including HSV-based cell segmentation and calculation. Stained breast cancer cells are segmented from the entire FOV using intensity thresholding. Single-cell morphological and apoptosis changes are characterized by calculating the circularity (c) and area (d) of each cell, respectively. The entire process is completed within 5 seconds.

The Auto-ICell image registration algorithm was employed to eliminate errors in the morphological and apoptotic analysis in single cells, under both bright-field and fluorescence-field imaging. In the bright-field context, sequential images of single cells encapsulated in droplets were synchronized using the image registration algorithm to offset disturbances caused by shifting, as depicted in **Fig. 4(a)**. During on-site experiments, a noticeable shifting of observed objects was recorded, which could potentially lead to detection inaccuracies. This shift could be attributed to a variety of factors, including thermal fluctuations, mechanical vibrations, and Brownian motion. The heat generated by the microscope's light source can cause temperature differentials in the medium, which in turn can generate convection currents capable of displacing cells or droplets. Likewise, changes in exposure due to operational settings can also result in detection errors. By employing the image registration and exposure averaging algorithm, the captured images were swiftly aligned within a span of 2 seconds, with the initial image acting as the reference flag. Similarly, in the fluorescent field as illustrated in **Fig. 4(b)**, sequential images of single cells encapsulated in droplets were synchronized using the same algorithm. The algorithm was further enhanced to incorporate additional analysis functions, such as HSV-based cell segmentation and calculation. Stained breast cancer cells were isolated from the FOV using intensity thresholding. To delineate the changes in single-cell morphology and apoptosis, the circularity (**Fig. 4(c)**) and area (**Fig. 4(d)**) of each cell were computed and graphically represented. These results offer a visual and quantitative depiction of the cell blebbing process.

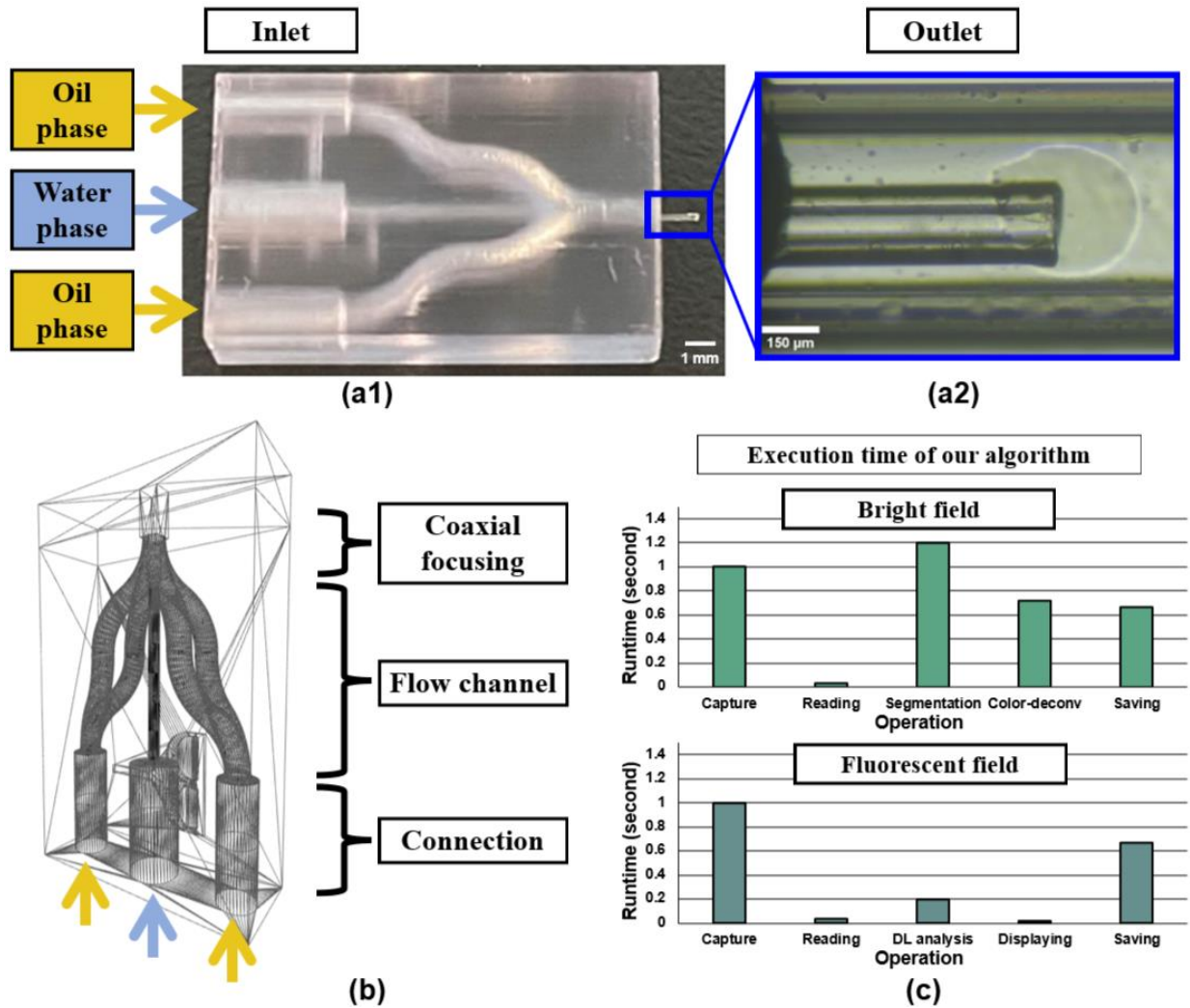

**Fig. 5 Characterization of the Auto-ICell system, including the customized 3D-printed droplet generation microfluidic chip and algorithm efficiency.** (a) Photographs of the 3D-printed microfluidic chip. (a1) The microfluidic chip is manufactured using a stereolithography 3D printer (resolution: 10 μm) with photosensitive resin as the printing material. The oil phase (continuous phase) and water phase (dispersed phase) are injected from the side and middle inlets, respectively, with the generated droplets collected from the outlet. Droplet size can be adjusted by fine-tuning the flow speed of the different phases. The assembly process incorporates glass capillaries into the 3D-printed chip to reduce surface roughness and increase transparency, thereby minimizing non-specific absorption and enhancing observation (Scale bar: 1 mm). (a2) The concentric configuration of inner and outer channels at the coflowing region with the assembled glass capillary is observable under a microscope (Scale bar: 150 μm). (b) Illustration of the 3D-printed droplet generator microfluidic chip microstructure. The chip features a connection section, flow channels, and a coaxial focusing section. The split winding subchannels are designed to generate substantial and balanced shear pressure around the dispersed phase. (c) Execution time of algorithms in bright field and fluorescent field. Processes include capture, reading, segmentation,

analysis, and display. The algorithms are capable of automatic image collection, analysis, displaying, and plotting with a delay of less than 1.772 seconds.

**Fig. 5 (a)** and **(b)** display photographs and microstructure illustrations of the 3D-printed droplet generation microfluidic chip, respectively. The chip consists of three sections: (1) the connection section, (2) the separating flow section, and (3) the converging flow section. The connection section facilitates the introduction of fluids through three cylindrical channels. The central channel, which carries the dispersed-phase fluid, has a diameter of 2.25 mm, while the other two channels on each side, which carry the continuous-phase fluid, have a diameter of 1.65 mm. Notably, the channels for the continuous-phase fluid bifurcate rapidly in the axial direction, with each channel dividing into two winding subchannels. These subchannels are designed to generate significant and balanced shear pressure around the dispersed phase in the 3D space at the co-flowing region. In the converging flow section, these subchannels converge with the central channel, resulting in the co-flowing of the dispersed-phase and continuous-phase fluids. The outlet, situated at the end of the generator, enables connection with other components such as glass capillaries and tubes for practical applications. The fabrication process involves model design, 3D printing, UV curing, and assembly with glass capillaries and tubing.

Our technique operates in two modes: offline and real-time, catering to user requirements. The offline mode allows for post-experiment analysis and parameter fine-tuning, while the real-time mode captures and displays one frame per second, involving four main computational operations. In terms of runtime, the offline analysis and display process takes approximately 0.274 seconds per FOV frame, enabling simultaneous analysis and result plotting. In the real-time mode, we capture and display one frame per second, which includes the following four computational operations: (1) reading images from the fluorescence microscope's API or SDK cable, taking around 0.033 seconds per image; (2) utilizing developed algorithms for image analysis. For bright fields, the major computational operations are depicted in **Fig.5 (c)**. This involves segmenting the reactor region from the full FOV image (8-bit, 1024 × 1024 pixels, 3 color channels) by identifying the background and applying binary masking (~1.198 s per image), performing immediate color deconvolution and plotting (~0.720 s), and saving the images on an internal solid-state drive (~0.668 s per image). For fluorescent fields, after capture and reading, deep-learning-enabled cell detection takes ~0.194 s, and displaying analyzed results of each detected cell on GUI takes ~0.004 s per image. In summary, the algorithms facilitate automatic image collection, analysis, display, and plotting with a delay under 1.772 seconds.

### 3. Discussion and Conclusion

Droplet microfluidics presents a promising avenue for investigating single cells or molecules, facilitating large-scale, high-throughput parallel screening processes, and precise control of interfacial properties. The development of a microfluidic device based on single-cell cultivation holds great potential for personalized medicine, offering advantages such as device miniaturization, significant reduction in reagent consumption, portability, low cost, and compatibility with traditional integrated circuit manufacturing processes. However, the progress of this field has been hindered by the high cost of facilities and the requirement for skilled manpower. To overcome these challenges, we have developed a 3D-printed microfluidic chip for rapid and cost-effective device fabrication, along with image-analysis algorithms for real-time automatic monitoring and quantification. We validated the Auto-ICell system through droplet reactor generation, color-based content analysis, and single-cell morphological and apoptotic analysis. By significantly reducing setup costs and enhancing operational performance, our Auto-ICell system demonstrates its superiority in improving the detection throughput of droplet single-cell analysis.

We compared our Auto-ICell system with other technologies in terms of device fabrication (**Table 1**) and operation processing (**Table 2**), respectively. Conventional methods for fabricating microfluidic chips, such as soft lithography and injection molding, are often characterized by high costs and time requirements. These limitations, including long cycle times and the need for expensive facilities, hinder the advancement of microfluidics, particularly in the early stages of research. Furthermore, the use of toxic chemicals (e.g., HF and SU8 solvent) in these methods contradicts the original objectives of microfluidics development, which strive for affordability, convenience, portability, environmental friendliness, and disposability[18]. Our prototyping method offers a cost-effective and time-saving solution for microfluidic chip fabrication, reducing the typical fabrication process from several days to less than 2 hours. Nevertheless, our Auto-ICell system offers substantial improvements in the detection throughput of droplet single-cell analysis, delivering notable benefits in terms of cost reduction and operational performance compared to conventional approaches. As the number of cells doubles with each successive generation and higher cell density increases the risk of confounding cell identities, the need for precision facilities and professional training becomes a barrier to ease of use. Additionally, conventional software often requires weeks to months for data processing.

The performance of our Auto-ICell system is influenced by various factors, including the resolution of the 3D-printed chip and image quality parameters such as focus, resolution, and luminance. To address these challenges, we are considering the utilization of a higher-resolution 3D printing machine or alternative printing materials to achieve higher resolution and reduced cycle time Additionally, we have optimized the imaging settings for both bright fields and fluorescent fields to ensure image sharpness, thereby enhancing the throughput and accuracy of the algorithm. Our experiments demonstrated that fluorescence noise, potentially caused by fouling or specks of dust, does not interfere significantly as it is much smaller than the detected objects. The Auto-ICell system employs disposable microfluidic chips for routine background signal evaluations. Future optimizations may involve further investigations into stain-less image analysis or virtual histological staining of biological samples for label-free and harmless diagnosis[33,34]. Furthermore, exploring the replacement of single-emulsion with multiple-emulsion[25,35] or all-aqueous emulsion[36,37] can be a fruitful area for further research as they offer improved biocompatibility.

In conclusion, the Auto-ICell system represents a significant step forward in the development of integrated automatic microfluidic systems for biochemical applications. It is the non-invasive, cost-effective, and user-friendly solution, providing robustness and time-saving capabilities. The insights gained from this study can contribute to the exploration of more advanced functionalities in current microfluidic systems and applications, including cell culture, biochemical microreactors, and drug carriers. Looking ahead, this paves the way for digital microfluidics to become a suitable platform for various applications, such as cell-based assays, synthetic biology, and point-of-care diagnostics[26,38–41].

**Table 1.** Comparison of Auto-ICell and conventional microfluidic chip fabrication techniques.

| | Microfluidic device fabrication techniques | | |
|---|---|---|---|
| | **Soft lithography** | **Injection molding** | **Our work** |
| Setup cost | ~$80k | >$50k | **<$10k** |
| Cost per print/materials | High | Low | **Low** |
| Cycle time | ~24h | 3 weeks | **<2h** |
| 3D capability | Layered 2D designs | Layered 2D designs | **3D designs** |
| Throughput | Low | Very high | **High** |

**Table 2.** Comparison of Auto-ICell and conventional droplet single-cell analysis techniques.

| | Single-cell droplet reactor observation techniques | | |
|---|---|---|---|
| | **Flow cytometry** | **Fluorescence microscope + ImageJ** | **Our work** |
| Setup cost | >$500k | <$10k | **<$10k** |
| Automaticity | High | Low | **High** |
| Real-time analysis capability | High | Low | **High** |
| Training for operators | Yes | Yes | **None** |
| Operation difficulty | High | Low | **Low** |

## 4. Materials and methods
### 4.1 3D-printed microfluidic chip and experimental setup

The customized microfluidic chip was designed using Autodesk Fusion 360 (Autodesk, Inc.). It consists of inlet, connection, and outlet sections. The connection part comprises three cylindrical channels: the central channel (2.25 mm in diameter) for the dispersed phase fluid, and the other two channels (1.65 mm in diameter) for the continuous phase fluid. To balance the shear force exerted by the continuous phase on the dispersed phase at the co-flowing region, the two continuous phase channels in the inset part are designed with bifurcated winding subchannels of 1 mm in diameter, which converge at the outlet section. The central dispersed phase channel has a diameter of 500 μm. The outlet part is a square channel with a length of 900 μm. The design file can be found in the Supporting Materials section. The 3D-printed microfluidic chip is manufactured utilizing a stereolithography 3D printer (resolution: 10 μm, Max X, Asiga, Alexandria, Australia) using photosensitive resin (DentaGUIDE, Asiga, Alexandria, Australia) as printing material. The assembly process is applied afterward as Fig.5(a) shows. A glass capillary with an inner diameter of 100 μm is placed in the central dispersed phase channel, which reduces the surface roughness of the channel to prevent non-specific absorption. In addition to this, a square glass capillary with a width of 500 μm is placed in the outlet section to keep coaxial with the central channel, allowing for the co-flowing. Furthermore, the increasing transparency of glass makes it possible to observe under the microscope and enable cell imaging.

The experimental platform is illustrated in **Fig. 1**, modified from our reported research works [20,42]. Monodisperse emulsions were generated using the 3D-printed microfluidic chip, with flow driven by two syringe pumps (LSP02-2A, Longer Precision Pump company) at the inlets. Mineral oil (Sigma-Aldrich, St. Louis, MO, USA) was used as the continuous-phase fluid, and the surfactant of span80 with 1% concertation was added to stabilize the droplets through the spontaneous accumulation at the droplet−carrier phase interface. Phosphate-buffered saline (PBS) with dye and suspension of stained cells were used as dispersed phases in different experiments respectively. Subsequently, the generated droplets were collected and transferred into a 96-well cell culture plate for cell cultivation and further observation.

### 4.2 Microreactors imaging and image processing

A high-speed camera (Mikrotron, Unterschleißheim, Germany) with a frequency of 1 kHz was employed to record the dynamic process of droplet generation via a 5× objective lens (Olympus, Tokyo, Japan). The imaging of the microreactors was performed using an inverted microscope (IX73, Olympus, Tokyo, Japan) coupled with a camera (CS126CU, Thorlabs, Newton, NJ, USA) in both brightfield and fluorescence fields. The double staining of cells was performed by using the fluorescence dyes Calcein-AM and propidium iodide (PI) following the manufacturer's instructions. PI ($\lambda_{ex}$ = 530 nm, $\lambda_{em}$ = 620 nm) is a nuclear staining reagent that can stain DNA and is often used to detect apoptosis since it can cross the damaged rather than the viable cell membrane. The dye Calcein-AM ($\lambda_{ex}$ = 490 nm, $\lambda_{em}$ = 515 nm) is a cell-staining reagent that can easily penetrate the membrane of living cells and thus fluorescently label living cells.

For brightfield image analysis, a custom color-deconvolution algorithm developed in MATLAB (MathWorks) was utilized. Sequential image frames were captured at a resolution of 1632×1232 pixels. For pre-processing the brightness was normalized, and the background was eliminated. This was accomplished by calculating a time-averaged image of the preceding static frames and subtracting it from the current raw image. The resulting background-subtracted full FOV image contained only the newly introduced droplets. The droplets were automatically detected and segmented from the full FOV for individual processing. The segmented image was then Gaussian-filtered, converted into a binary image, and filtered based on desired criteria, including color distribution. To reduce misdetection events caused by sensor noise, binary contours with an area of a few pixels were removed. The resulting binary contours represented the shapes and locations of the droplets appearing in the FOV, and their morphological information was used to filter each contour based on desired criteria, such as color distribution and content concentrations. The segmented image was further analyzed to detect and plot the intensity of target colors. Color calibration was performed using the RGB values of droplets containing only one dye. Rigid image registration was carried out to correct for lateral specimen displacements, aligning each frame to a reference image and aligning different video sequences from each experiment to each other. Lateral motion corrections were typically below 1 μm.

Monitoring for multiple fluorescent channels was initially performed by transferring the aforementioned algorithms into the hue, saturation, value (HSV) model. By setting specific thresholds, cells with strong fluorescence intensities were segmented and further analyzed to extract information such as area, perimeter, and circularity. To optimize the results, a custom deep-learning-enabled algorithm was developed in Python (The Python Software Foundation). This algorithm enabled fluorescent color transfer from different colors by rewriting the RGB values of different RGB channels. Additionally, a GUI was developed by utilizing Cellpose[32] for automatic detection. Subsequently, the masks of segmented images were analyzed to obtain information including area, perimeter, and circularity.

### 4.3 Cell culture

MDA-MB-231 cells were obtained from the American Type Culture Collection (ATCC, Wuhan, Wuhan Procell Life Technology Company) and cultured in Roswell Park Memorial Institute 1640 medium supplemented with 10% fetal bovine serum (FBS, FBS, Gibco, USA). The cells were seeded in a tissue culture plate as a 2D monolayer. Subsequently, MDA-MB-231 cells ($2\times10^6$ cells/mL) were seeded in a microwell plate to form spheroids and maintained at 37°C in a 5% carbon dioxide ($CO_2$) environment. The culture medium was replaced every two days, and the spheroids were cultured for up to 14 days.

### 4.4 Cell apoptotic analysis by Calcein acetoxymethyl ester/propidium iodide (Calcein-AM/PI) staining

The viability of cancer cells was evaluated using the Calcein-AM/PI Double Stain Kit (Dojindo Laboratories, Japan). This kit enables simultaneous staining of both living and dead cells using two fluorescent dyes. Living cells produce green fluorescence due to the conversion of Calcein acetoxymethyl ester (Calcein-AM) to Calcein, while dead cells produce red fluorescence due to the presence of propidium iodide (PI). Cancer cells were seeded at $8 \times 10^3$ cells/well in 96-well plates, as described above. The cells were stained with the Calcein-AM/PI Kit following the manufacturer's instructions. Subsequently, $5\times10^5$ cells were collected and washed twice with precooled PBS. The cells were then resuspended in 100 μL of binding buffer. The suspended cells were injected into a customized microfluidic system for reactor generation and further analysis. All the above procedures were performed gently in the dark at room temperature. Finally, living cells (green cytoplasmic fluorescence) and dead cells (red nucleus fluorescence) were observed using an inverted fluorescence microscope for image analysis to measure apoptosis.

### 4.5 Statistical Analysis

Statistical analysis was performed using GraphPad Prism software (GraphPad Software). Data are presented as mean ± standard deviation (SD) with n ≥ 3. Statistical comparisons between different groups were conducted using one-way analysis of variance (ANOVA). Tukey's multiple comparison test was used to evaluate statistical differences between samples. A significance level of $p < 0.05$ was considered for all analyses.

## 6. Acknowledgements

The authors are grateful to the funding support from Guangdong Basic and Applied Basic Foundation (2020A1515111053 and 2022A1515011566), the Hong Kong Research Grants Council (project reference: GRF14204621, GRF14207121, GRF14207920, GRF14207419, GRF14203919, N_CUHK407/16), the Marine Conservation Enhancement Fund (MCEF20108_L02), and the Innovation and Technology Commission (project reference: GHX-004-18SZ).

The authors would like to acknowledge Mr. Yuankai Ma (Department of Mechanical Engineering, Tsinghua University), Dr. Nawaz Mehmood, and Ms. Syeda Aimen Abbasi (Department of Biomedical Engineering, The Chinese University of Hong Kong) for their support in the project development.

## 7. Author information

Authors and Affiliations

**Department of Biomedical Engineering, The Chinese University of Hong Kong, Shatin, Hong Kong SAR, 999077, China**

Yuanyuan Wei, Sai Mu Dalike Abaxi, Guangyao Cheng, Dehua Hu, Yi-Ping Ho, Wu Yuan & Ho-Pui Ho

**Department of Biomedical Engineering, Shantou University, Shantou, China**

Meiai Lin & Shanhang Luo

**Department of Biomedical Engineering, National University of Singapore, Singapore**

Shanhang Luo

**Department of Electrical Engineering, Tsinghua University, Beijing, China**

Liwei Tan

**Electrical and Computer Engineering Department, University of California, Los Angeles, California, 90095, USA**

Bijie Bai

**Centre for Biomaterials, The Chinese University of Hong Kong, Hong Kong SAR, 999077, China**

Yi-Ping Ho

**Hong Kong Branch of CAS Center for Excellence in Animal Evolution and Genetics, Hong Kong SAR, 999077, China**

Yi-Ping Ho

**State Key Laboratory of Marine Pollution, City University of Hong Kong, Hong Kong SAR, 999077, China**

Yi-Ping Ho

Contributions

Y. Wei, S. Abbasi, and M. Lin contributed to the study's conception and design. M. Lin developed the microfluidic system. B. Bai and Y. Wei developed the color-based image analysis algorithm. Y. Wei and S. Abbasi developed the deep-learning-enabled multiple fluorescent channels analysis algorithm. M. Lin and S. Luo performed microfluidic experiments. Y. Wei, S. Luo, S. Abbasi, and L. Tan conducted the image analysis experiments and data analysis. Y. Wei wrote the manuscript with contributions from all authors. W. Yuan and H. Ho conceived the project and supervised the research.

Corresponding author

Correspondence to Wu Yuan and Ho-Pui Ho.

8. Ethics declarations

No conflict of interest

9. Additional information

Electronic supplementary material

Supplementary Information